\documentclass[%
pre, %
twocolumn,%
secnumarabic,%
showpacs, %
amssymb, amsmath, nobibnotes, aps]{revtex4}
\usepackage{docs}%
\usepackage{bm}%

\usepackage{subfigure}
\usepackage{epsfig}
\usepackage{latexsym}
\usepackage{amsthm}
\usepackage{amsfonts}
\usepackage{amsbsy}
\usepackage{here}
\usepackage{psfrag}
\usepackage{verbatimfiles}

\expandafter\ifx\csname package@font\endcsname\relax\else
 \expandafter\expandafter
 \expandafter\usepackage
 \expandafter\expandafter
 \expandafter{\csname package@font\endcsname}%
\fi

\begin{document}

\title{Energy funneling in a bent chain of Morse oscillators with long--range 
coupling}%

\author{P.V. Larsen}%
\email{pvl@imm.dtu.dk}
\author{P.L. Christiansen}%
\affiliation{Informatics and Mathematical Modelling, 
             Technical University of Denmark, DK-2800 Kgs. Lyngby, Denmark}

\author{O. Bang}%
\affiliation{Informatics and Mathematical Modelling, 
		Technical University of Denmark, DK-2800 Kgs. Lyngby, Denmark}
\affiliation{Research Center COM, Technical University of Denmark, 
		DK-2800 Kgs. Lyngby, Denmark}

\author{J.F.R. Archilla}
\affiliation{Departamento de Fisica Aplicada I, Universidad de Sevilla, 
		Avda. Reina Mercedes s/n, 41012 Sevilla, Spain}

\author{Yu.B. Gaididei}%
\affiliation{Bogolyubov Institute for Theoretical Physics, 03143 Kiev, Ukraine}


\begin{abstract}
A bent chain of coupled Morse oscillators with long--range 
dispersive interaction is considered. Moving localized excitations may be 
trapped in the bending region. Thus chain geometry acts like an impurity. An 
energy funneling effect is observed in the case of random initial conditions.
\end{abstract}

\maketitle

\pagebreak

\section{\label{sec:Intro}Introduction}

Nonlinear excitations (\emph{solitons}, \emph{discrete breathers}, 
\emph{intrinsic localized modes}, etc.) have been drawing increasing attention 
over recent years and are widely believed to be responsible for several 
effects in molecular chains, such as charge and thermal conductivity, energy 
transfer and localization, etc. (see reviews in 
\cite{Davy, PR 217, Yaku, NLinBio, PR 295, RMP 60} {\it e.g.}).\par 
Initially, the geometrical features of the polymers and biopolymers were 
essentially neglected and energy transfer and localization was mostly 
attributed to inhomogeneities and impurities 
\cite{PRE 53/1, PRE 67, PRE 55/4, JPA 35_10519, Muto, PRE 49} or nonlinear 
excitations \cite{PLA 154, Bang/Peyrard, PRE 55/4, JBP 25, PRL 70, PRE 51}. 
Also, discreteness plays an important role for the localization of these 
excitations. 
The inhomogeneities have been modelled by different masses at various chain 
sites \cite{PRE 49, PRE 53/1, PRE 67}, by changes in the coupling between 
molecular sites \cite{PRE 55/4, PRE 67} or by different on--site potentials 
\cite{JPA 35_10519} as well as conformational defects \cite{PRE 51}. 
In general, impurities have been shown to act as filters governing the 
progression of incoming excitations. Thus both reflection, trapping and 
transmission of incident moving discrete breathers through the impurity region 
can occur \cite{JPA 35_10519, PRE 49, PRE 67, PRE 51, PRE 53/1, PRE 55/4}. 
Similar effects have been observed through collisions between moving discrete 
breathers, thus \cite{PRE 55/4, Bang/Peyrard, PRL 70} showed how  stationary 
large amplitude discrete breathers, on the average, absorb energy from 
colliding breathers of smaller amplitude. Thus the large amplitude breather 
may play a role similar to that of an impurity \cite{PRE 49}.\par
Recently, both long--range dipole--dipole interaction \cite{PD 163, PLA 249}, 
helicity \cite{PLA 253, JBP 24} and curvature 
\cite{EPL 59, JPA 34_8465, Cond 13} have been included in the nonlinear 
transport theory, as well as combinations of these effects 
\cite{JPA 35_8885, PRE 66, JPA 34_6363, PRE 62, PLA 299}. It has been shown 
that chain geometry induces effects similar to those of impurities 
\cite{EPL 59, JPA 34_8465, JPA 35_8885, Cond 13}.\par
Special attention has been paid to models of biological macromolecules, such as 
proteins \cite{Davy, PR 217, PLA 154} and DNA 
\cite{PRE 53/1, JPA 35_10519, PLA 154, Davy, PD 57, JBP 25, Muto}. These are 
obvious choices for more complex geometric models, as their structure is 
crucial for their functionality \cite{Und DNA, Saenger}. A widely used model 
was presented by Peyrard and Bishop \cite{PRL 62} in the context of 
statistical mechanics. In biological environments, thermal fluctuations are 
always present and have been considered in 
\cite{PD 119, PRE 60, Muto, PRE 47/1, PD 57}, {\it e.g.} In these refs. it was 
shown that solitons or discrete breathers can be generated from initial random 
thermal fluctuations.\par
The aim of the present work is to study the interplay between chain 
geo\-me\-try and long--range in\-ter\-action in an augmented Peyrard--Bishop 
model. We show how a new mechanism for energy accumulation in the system --- 
\emph{funneling} --- may be provided by the geometry of the chain. We consider 
a simple approximative description of the long--range intersite coupling by 
modelling it as a dipole--dipole--like interaction. Using an attractive 
long--range interaction, we study the effect of the geometry of the bent chain 
on the dynamics. The particular shape of the bend turns out to make no 
qualitative difference in terms of trapping and funneling. We therefore choose 
a simple wedge--shaped geometry. As initial conditions we use discrete 
breathers as well as randomly distributed fluctuations 
\cite{Bang/Peyrard, Muto}.\par
In Sec.~\ref{sec:model2} we introduce the model, which  includes the 
long--range interactions and the chain geometry. In Sec.~\ref{sec:results}, we 
investigate breather dynamics in the system and in Sec.~\ref{sec:random} 
random initial conditions are considered. Sec.~\ref{sec:conclu} summarizes our 
results and contains a discussion.

\section{\label{sec:model2}The Model}
\begin{figure}[h] 
    \scriptsize
      \psfrag{y2}[br][br]{$1$}
      \psfrag{y1}[br][br]{$0$}
      \psfrag{x1}[cc][cc]{$0$}
      \psfrag{x2}[cc][cc]{$4$}
      \psfrag{x3}[cc][cc]{$8$}
      \psfrag{Morse potential}[tt][tc]{Potential, $V(u_n)$}
      \psfrag{Displacement}[Bc][Bc]{Displacement, $u_n$}
    \normalsize
  \centerline{
    \epsfig{file=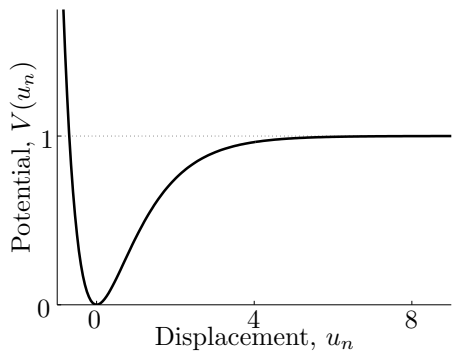, width=6 cm, angle=0} } 
  \caption{The Morse potential, $V(u_n)=[\exp(-u_n)-1]^2$.}
  \label{fig:Morse}
\end{figure}

We consider a one dimensional lattice of Morse oscillators with the 
Hamiltonian density 
\begin{eqnarray}
\mathcal{H}_n &=&  \frac{1}{2} \dot{u}_n^2  + 
   \frac{C}{2} \left( u_n - u_{n-1} \right)^2 \nonumber\\
  && + \left[ \exp (-u_n) -1 \right]^2 -
   \frac{1}{2} \displaystyle \left. \sum_{m} \right.^{\prime} J_{nm} u_n u_m, 
   \label{eq:model2}
\end{eqnarray}
\noindent where prime indicates $m \neq n$ in the summation. The Hamiltonian 
for the system becomes $H = \sum_{n=-N}^{n=N} \mathcal{H}_n$, where the total 
number of sites is $N_T = 2N+1$. In Eq.~(\ref{eq:model2}) the first term is 
the kinetic energy at the $n$'th site. Then  follows a harmonic potential 
interaction between neighboring sites, $C$ being the dispersion parameter. An 
on--site Morse potential (shown in Fig.~\ref{fig:Morse}) describes the atomic 
interaction. Finally, there is a summation of long-range interactions in which 
the coefficients are given by 
\begin{equation} \label{eq:J2}
J_{nm} = \frac{J_0}{\left| {\bm{r}}_n - {\bm{r}}_m \right|^3},
\end{equation}
\noindent where $J_0$ is a strength parameter, and $\mathbf{r}_n$ denotes the 
position of the n'th site. In our model the distance between neighboring 
sites, $\left| {\bm{r}}_{n+1} - {\bm{r}}_{n} \right|$, is constant and 
normalized to unity.\par
\begin{figure}[h]
 \large
  \psfrag{xn}[cl][cc]{$x_n$} 
  \psfrag{yn}[cc][cc]{$y_n$} 
  \psfrag{Th}[cc][cc]{{\Large $\theta$}} 
  \psfrag{1}[cc][cc]{$1$} 
  \psfrag{n=-N}[cr][cr]{$n=-N$} 
  \psfrag{n=-1}[br][cr]{$n=-1$} 
  \psfrag{n=0}[cc][cc]{$n=0$} 
  \psfrag{n=1}[bl][cl]{$n=1$} 
  \psfrag{n=N}[cl][cl]{$n=N$} 
 \normalsize
  \includegraphics[width=6cm, height=4cm]{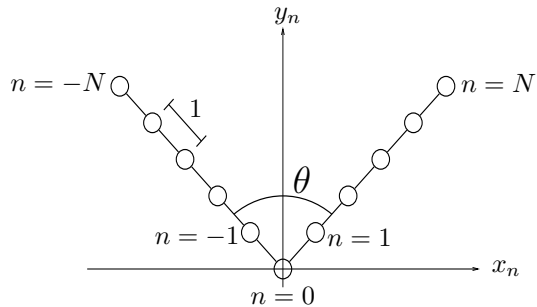}
 \caption{\label{fig:wedgechain}Wedge chain with opening angle, $\theta$.}
\end{figure}

From the Hamiltonian we get the equations of motion
\begin{eqnarray}
  & \ddot{u}_n + C \left( 2u_n - u_{n-1} - u_{n+1} \right) & \nonumber \\
 & - 2 e^{-u_n} \left[ e^{-u_n} - 1 \right] - 
 \displaystyle \left.\sum_{m} \right.^{\prime} J_{nm} u_m = 0 &. 
 \label{eq:moteqs}
\end{eqnarray}
\indent At the ends of the molecule we use the free boundary conditions
\vspace{-3mm}
\begin{eqnarray}
	u_{-N-1} & = & u_{-N},\\
	u_{N+1} & = & u_{N}.
\end{eqnarray}
\indent The wedge-shaped chain is given by 
\begin{displaymath}
\bm{r}_n = \left( x_n, y_n \right)= (n \sin \frac{\theta}{2}, 
            |n| \cos \frac{\theta}{2}),
\end{displaymath}

\noindent where $\theta$ denotes the fixed wedge angle (see 
Fig.~\ref{fig:wedgechain}). We note that the geometry of the chain only comes 
into play through the long--range interactions. In fact, earlier studies of 
the long--range effect in curved molecular chains show that the exact form of 
the  additional dispersion is not crucial as long as it decreases rapidly with 
distance \cite{JBP 25, PRE 62}.\par
Throughout the paper we use $C=0.075$ (a typical value for DNA 
\cite{PD 163, PRE 47/1, PLA 299}) and $J_0 = 0.5$ (based on estimates of the 
dipole moment \cite{PD 163}).\par
\begin{figure}[ht]
  \centerline{
    \footnotesize
      \psfrag{I}[bc]{{\scriptsize $m\!+\!1$}}
      \psfrag{II}[c]{}
      \psfrag{III}[bc]{{\scriptsize $m\!+\!\!5$}}
      \psfrag{IV}[c]{}
      \psfrag{V}[bc]{{\scriptsize $m\!+\!9$}}
      \psfrag{0}[Br]{$0$}
      \psfrag{1}[Br][Br]{$J_0$}
    \normalsize
      \psfrag{(a)}[cc]{{\large \hspace{0mm}(a)}}
      \psfrag{(b)}[cc]{{\large \hspace{0mm}(b)}}
      \psfrag{YlabelI}[cc][tc]{{$J_{n,-3}$}}
      \psfrag{YlabelII}[cc][tc]{{$J_{n,-2}$}}
      \psfrag{Site}[c]{Site number, $n$}
    \includegraphics[width=4 cm, height=6cm, angle=0]{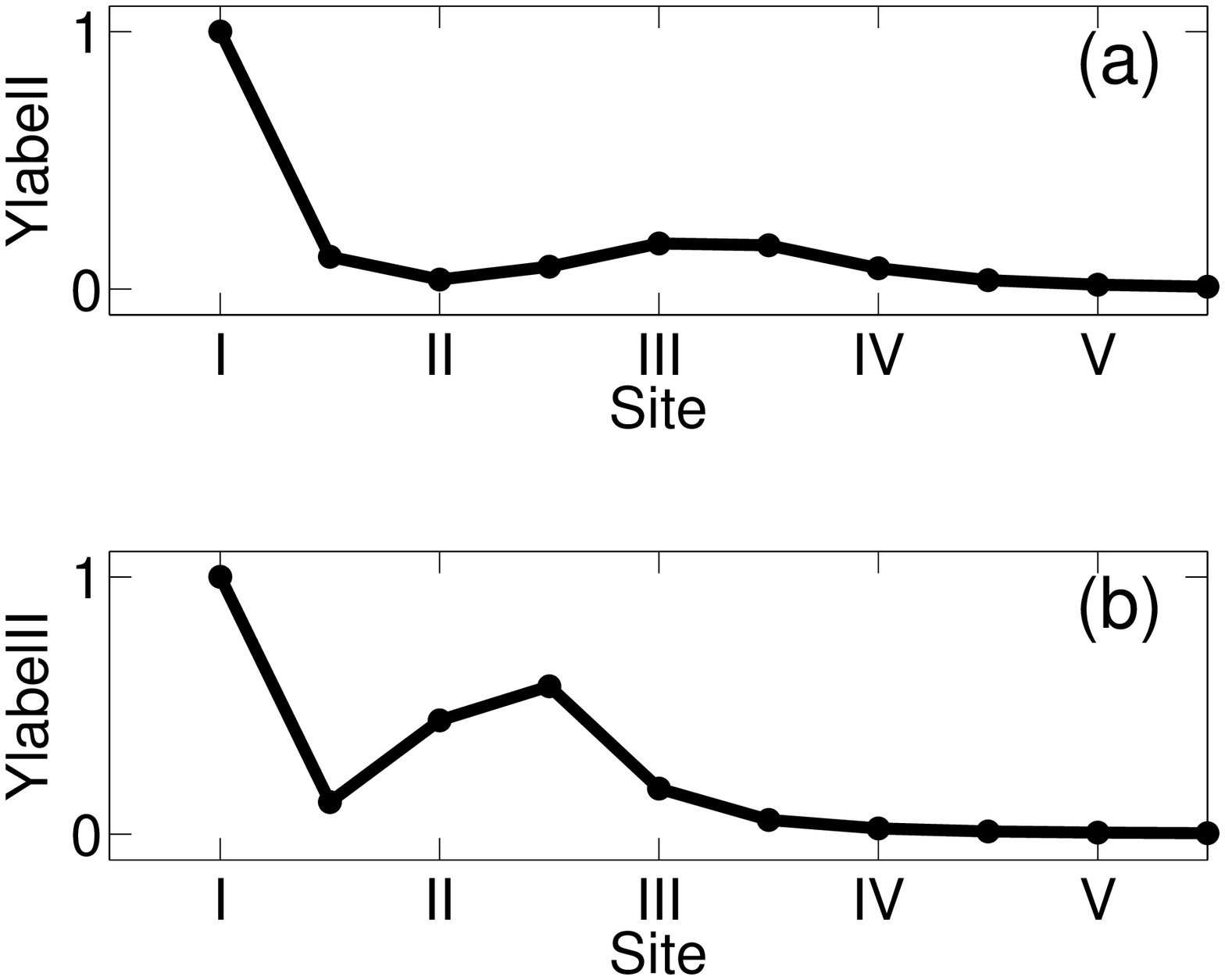}
  	\hspace{1cm}
      \psfrag{n=0}[cc][cc]{$n=0$}
      \psfrag{n=m=-3}[cr][cr]{{\scriptsize $n=m=-3\,$}}
      \psfrag{n=m=-2}[cr][cr]{{\scriptsize $n=m=-2\,$}}
    \includegraphics[width=4 cm, height=6cm, angle=0]{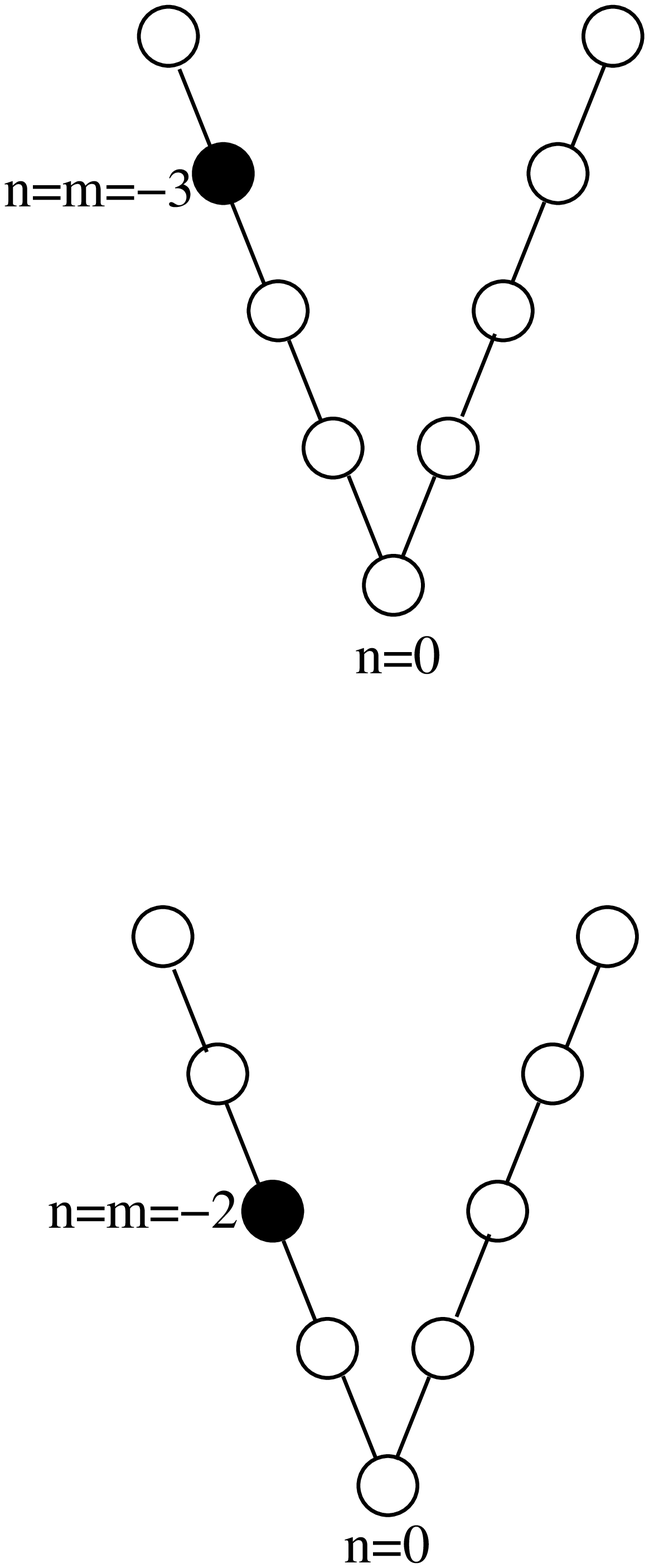}}
  \caption{Long--range interaction coefficients $J_{nm}$ for fixed (a) $m=-3$, 
  (b) $m=-2$. $\theta=35^{\circ}$.}
  \label{fig:Jterm}
\end{figure}

 
Fig.~\ref{fig:Jterm} gives a detailed picture of the long--range interaction 
by plotting the interaction coefficient $J_{nm}$ versus $n$, for fixed (a) 
$m=-3$ and (b) $m=-2$. We see that the closer $m$ gets to the bend at $n=0$, 
the higher the shoulder in the $J_{nm}$--profile becomes. This feature 
suggests an analogy in which the bend acts as an impurity. Away from the bend 
this effect rapidly drops off.\par
\begin{figure}[h] 
      \psfrag{VEff}[cc][cc]{$V_n^{\mathrm{Eff}}$}
      \psfrag{-1.2}[cr][cr]{$-1.2$}
      \psfrag{-1.3}[cr][cr]{$-1.3$}
      \psfrag{-1.25}[cr][cr]{$$}
      \psfrag{-1.35}[cr][cr]{$$}
      \psfrag{-15}[cc][cc]{$-15$}
      \psfrag{-10}[cc][cc]{$-10$}
      \psfrag{-5}[cc][cc]{$-5$}
      \psfrag{0}[cc][cc]{$0$}
      \psfrag{5}[cc][cc]{$5$}
      \psfrag{10}[cc][cc]{$10$}
      \psfrag{15}[cc][cc]{$15$}
      \psfrag{Site}[cc][cc]{Site, $n$}
  \centerline{
      \epsfig{file=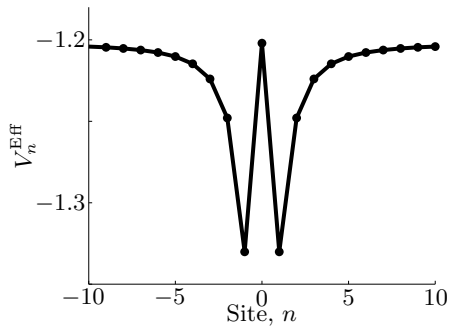, width=6 cm, angle=0} } 
  \caption{Effective potential, 
        $V_{n}^{\mathrm{Eff}} \equiv - \sum_{m}^{\prime} J_{nm}$, versus 
	site number, $n$.}
  \label{fig:IC}
\end{figure}

The long--range interaction in curved chains may be represented as follows: 
\begin{eqnarray*}
  - \sum_n \left. \sum_{m} \right.^{\prime} J_{nm} u_n u_m & = 
  & \frac{1}{2} \sum_n \left. \sum_{m} 
     \right.^{\prime} J_{nm} \left( u_n - u_m \right)^2 \\ 
  & & \quad + \sum_n V^{\mathrm{Eff}}_{n} u_n^{2},
\end{eqnarray*}
where the first summations on the right hand side correspond to the 
inhomogeneous dispersion (seen in Fig~\ref{fig:Jterm}), while the second 
summation, in which $V_{n}^{\mathrm{Eff}} \equiv - \sum_{m}^{\prime} J_{nm}$ 
is introduced, corresponds to an \emph{effective on--site potential} 
\cite{PRE 62}. The potential $V_n^{\mathrm{Eff}}$ has the double--well 
profile, shown in Fig.~\ref{fig:IC}.\par
In the following we shall see how the impurity, or effective on--site 
potential, introduced by the bend can reflect, trap or transmit incoming 
excitations.

\section{\label{sec:results}Energy trapping}
In this Section we consider the interaction between the bend and the incoming 
localized excitations.\par
A 4th order Runge--Kutta solver is used to simulate Eq.~(\ref{eq:model2}) on a 
chain with $N_T=301$ sites. A stepsize in time of 0.005 ensures conservation 
of the Hamiltonian to a relative accuracy of $10^{-10}$ throughout. We use a 
Gaussian initial condition
\begin{equation} \label{eq:gauss}
u_n(t) = A \exp \left[ -k \big( (n-\nu)-vt \big)^2 \right],
\end{equation}
\noindent where site $\nu$ denotes the initial position of the center of mass.
\par
In the following we use the velocity $v=0.2$, the width $k=0.2$ and the 
amplitude $A=0.5$, because they turn out to provide the right balance between 
nonlinearity and dispersion to allow the initial condition to evolve rapidly 
into a discrete moving breather. Insertion of Eq.~(\ref{eq:gauss}) with these 
parameter values into the total Hamiltonian gives $H=0.13$. After some initial 
radiation, the moving breather turns out to possess the energy, 
$H_s \approx 0.08$.\par
In the following sections we present the numerical simulations of the chain 
dynamics.\par
\subsection{Breather dynamics}
\begin{figure}[h] 
  \centerline{
      \psfrag{y5}[Br][Br]{{\scriptsize $800$}}
      \psfrag{y4}[Br][Br]{{\scriptsize $600$}}
      \psfrag{y3}[Br][Br]{{\scriptsize $400$}}
      \psfrag{y2}[Br][Br]{{\scriptsize $200$}}
      \psfrag{y1}[Br][Br]{{\scriptsize $0$}}
      \psfrag{x1}[cc][cc]{{\scriptsize $-100$}}
      \psfrag{x2}[cc][cc]{}
      \psfrag{x3}[cc][cc]{{\scriptsize $0$}}
      \psfrag{x4}[cc][cc]{}
      \psfrag{x5}[cc][cc]{{\scriptsize $100$}}
      \psfrag{Time, t}[Bc][tc]{Time, $t$}
      \psfrag{Site, n}[cc][bc]{Site, $n$}
      \psfrag{(a)}[cl]{{\large \hspace{0mm}(a)}}
      \psfrag{(b)}[cl]{{\large \hspace{0mm}(b)}}
     \epsfig{file=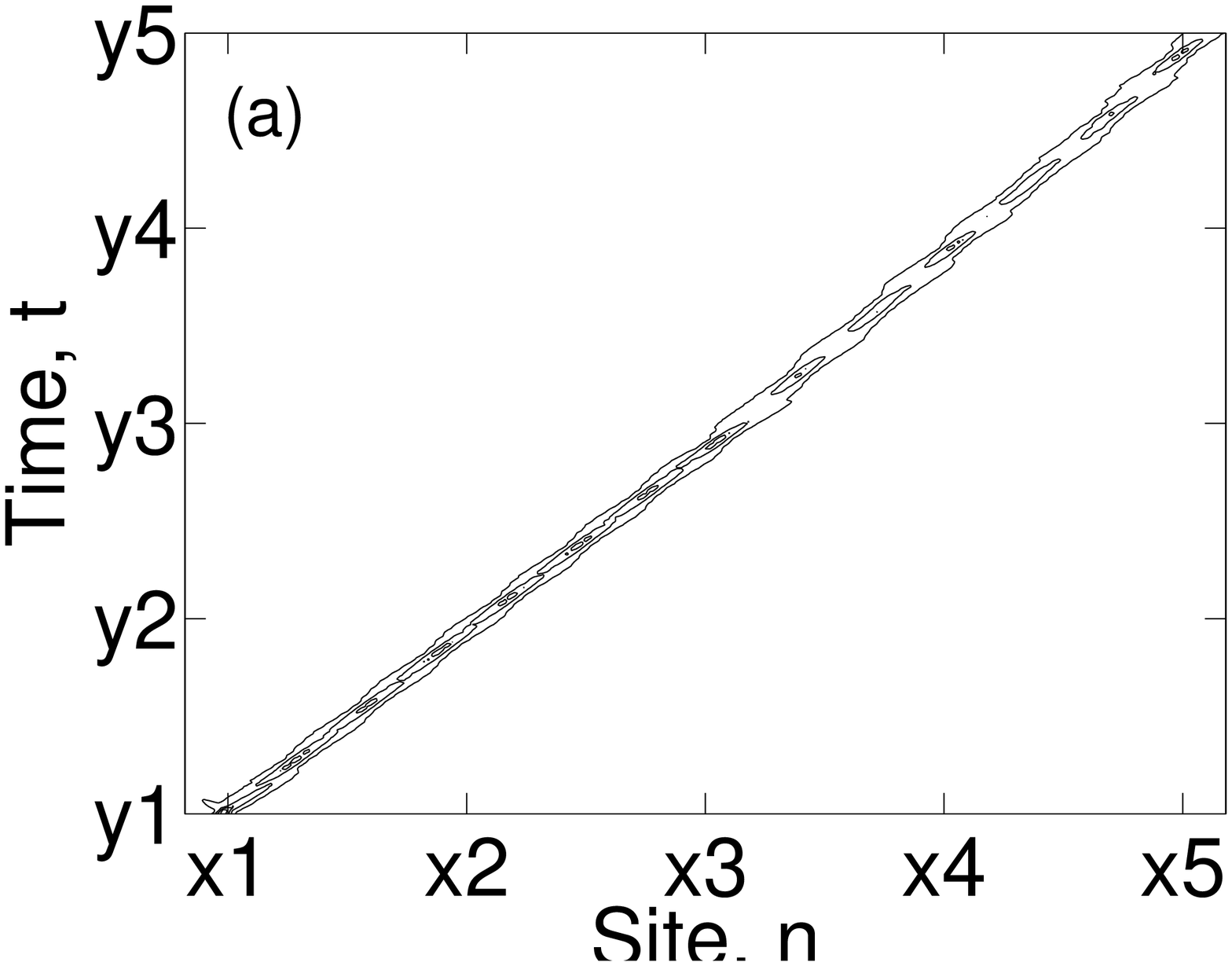, width=4 cm, angle=0}
      \hspace{2mm}
     \epsfig{file=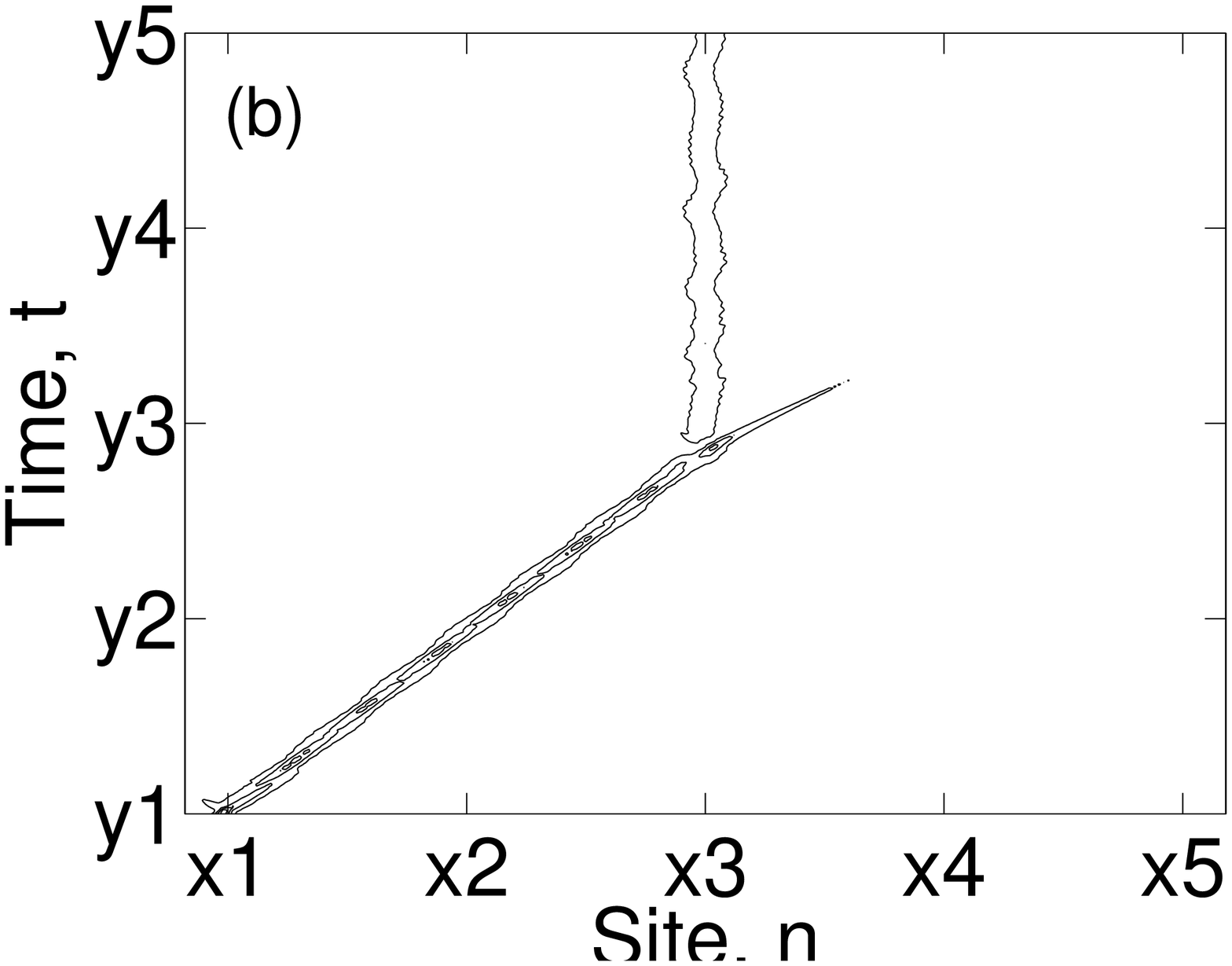, width=4 cm, angle=0} }
  \caption{Contour plots for the evolution of the Hamiltonian density, 
  	$\mathcal{H}_n$, for $N_T=301$, $J_0=0.5$, $v=0.2$, $k=0.2$, $A=0.5$ 
	and $\nu=-100$, yielding $H=0.13$. 5 equidistant contours 
	$H=0.005 \ldots 0.05$. (a) $\theta=140^{\circ}$ (transmission), (b) 
	$\theta=95^{\circ}$ (trapping).}
  \label{fig:contour}
\end{figure}

In Fig.~\ref{fig:contour} we  show contour plots for the evolution of the 
Hamiltonian density. A weak bend with $\theta=140^{\circ}$ has no noticeable 
effect on the breather (Fig.~\ref{fig:contour}a) and only a slight decrease of 
the velocity after passage of the center region is observed. In contrast for a 
stronger bend, $\theta=95^{\circ}$, a considerable part of the excitation is 
trapped at the tip of the wedge $n=0$.\par
Very strong bends (smaller wedge angles) turn out to result in reflection of 
the incident breathers. Such scattering properties and their dependence on the 
strength of the bend--induced impurity, are similar to those of a linear 
impurity \cite{PRE 63} and those of large amplitude breathers acting as an 
effective impurity \cite{Bang/Peyrard}. We stress that the specific shape of 
the bend does not affect the scattering properties of the bent chain. Thus 
similar properties were observed in a parabolic chain \cite{PVL}.\par
\begin{figure}[ht] 
  \centerline{  
      \psfrag{y3}[cr][cr]{$1$}
      \psfrag{y2}[cr][cr]{}
      \psfrag{y1}[cr][cr]{$0$}
      \psfrag{x1}[cc][cc]{$0$}
      \psfrag{x2}[cc][cc]{$200$}
      \psfrag{x3}[cc][cc]{$400$}
      \psfrag{x4}[cc][cc]{$600$}
      \psfrag{x5}[cc][cc]{$800$}
      \psfrag{HcH}[tc][tc]{$H_c/H_s$}
      \psfrag{Time, t}[tc][bc]{Time, $t$}
      \psfrag{(a)}[cl]{{\large \hspace{0mm}(a)}}
      \psfrag{(b)}[cl]{{\large \hspace{0mm}(b)}}
     \epsfig{file=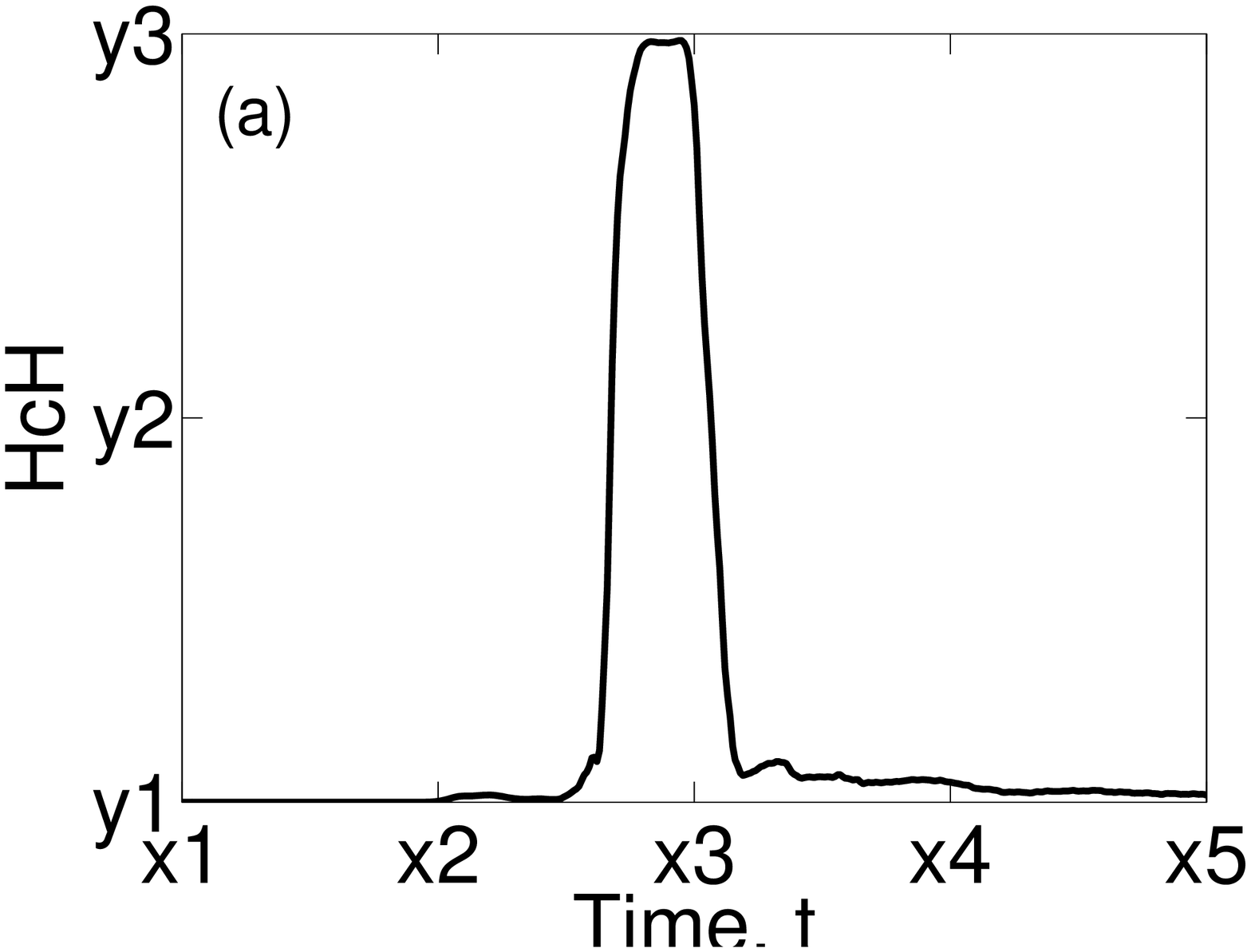, width=4 cm, angle=0} 
      \hspace{2mm}
     \epsfig{file=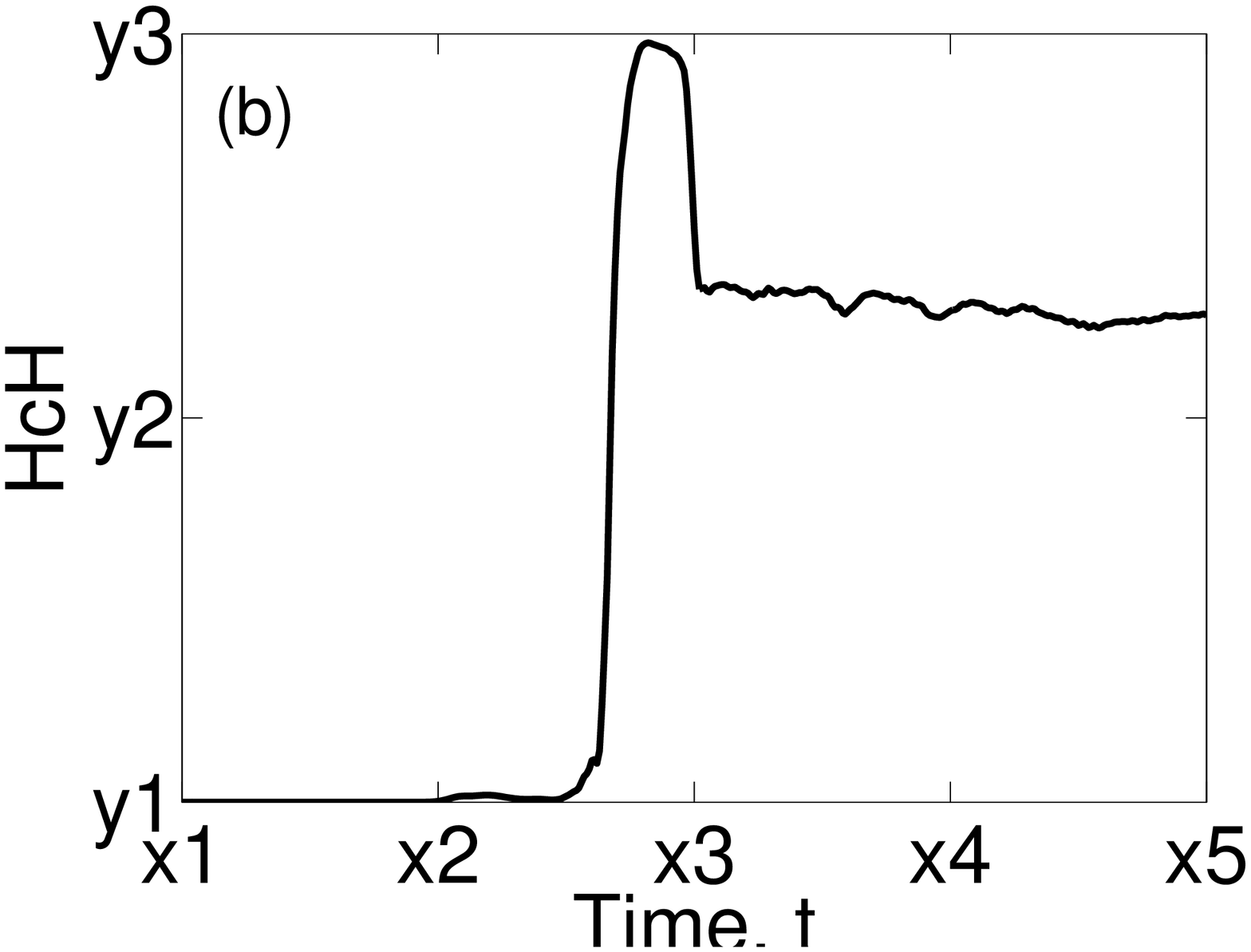, width=4 cm, angle=0}}
  \caption{Relative central energy, $H_c/H_s$, versus time for the simulations 
  	in Fig.~\ref{fig:contour}. (a) $\theta=140^{\circ}$, (b) 
	$\theta=95^{\circ}$.}
  \label{fig:Hcen}
\end{figure}

To analyze the processes in detail, we calculate the central energy, $H_c$, in 
21 sites around $n=0$ (21 being a typical span of the denaturation bubble of 
the DNA molecule \cite{JBP 24, PRE 47/1}) 
\begin{equation} \label{eq:Hcen}
	H_{\textrm{c}} = \sum_{n=-10}^{10} \mathcal{H}_n,
\end{equation}
\noindent where $\mathcal{H}_n$ is given by Eq.~(\ref{eq:model2}). The results 
are shown in Fig.~\ref{fig:Hcen}. In the transmission case with a small bend, 
Fig.~\ref{fig:Hcen}(a), nearly all the energy leaves the region. In the 
trapping case with a stronger bend, Fig.~\ref{fig:Hcen}(b), the trapped energy 
is stabilized at about $H_{c}/H_s \approx 64\%$. Thus only part of the energy 
is trapped.\par 
\begin{figure}[h] 
  \centerline{
      \psfrag{y8}[cr][cr]{}
      \psfrag{y4}[cr][cr]{$0.6$}
      \psfrag{y7}[cr][cr]{}
      \psfrag{y3}[cr][cr]{$0.4$}
      \psfrag{y6}[cr][cr]{}
      \psfrag{y2}[cr][cr]{$0.2$}
      \psfrag{y5}[cr][cr]{}
      \psfrag{y1}[br][cr]{$0$}
      \psfrag{x1}[cc][cc]{$80$}
      \psfrag{x2}[cc][cc]{$90$}
      \psfrag{x3}[cc][cc]{$100$}
      \psfrag{x4}[cc][cc]{$110$}
      \psfrag{x5}[cc][cc]{$120$}
      \psfrag{HcH}[bc][tc]{$H_c/H_s$}
      \psfrag{Wedge angle}[tc][cc]{Wedge angle, $\theta$}
    \epsfig{file=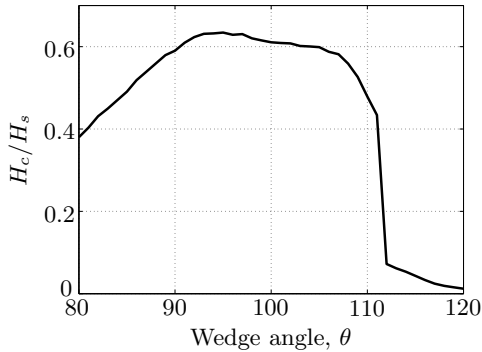, width=6 cm, angle=0} } 
  \caption{Relative central energy, $H_c/H_s$, at $t=800$ versus wedge angle, 
  	$\theta$. System parameters as in Fig.~\ref{fig:contour}.}
  \label{fig:max_Hcen_theta}
\end{figure}

The trapped energy portion, $H_c/H_s$, calculated at time = 800 is shown in 
Fig.~\ref{fig:max_Hcen_theta} as a function of the wedge angle, $\theta$. For 
smaller and larger $\theta$--values, the energy is lost through reflection and 
transmission, respectively. Effecient trapping is found for intermediate wedge 
angles, $90^{\circ} < \theta < 107^{\circ}$. The optimal wedge angle for 
trapping is seen to be around $\theta=95^{\circ}$.\par
\section{\label{sec:addsol}Multiple breather dynamics}
\begin{figure}[h] 
  \centerline{
      \psfrag{y5}[cr][cr]{$\phantom{-}100$}
      \psfrag{y4}[cr][cr]{$\phantom{-}50$}
      \psfrag{y3}[cr][cr]{$0$}
      \psfrag{y2}[cr][cr]{$-50$}
      \psfrag{y1}[cr][cr]{$-100$}
      \psfrag{x1}[cc][cc]{$0$}
      \psfrag{x2}[cc][cc]{$500$}
      \psfrag{x3}[cc][cc]{$1000$}
      \psfrag{x4}[cc][cc]{$1500$}
      \psfrag{x5}[cc][cc]{$2000$}
      \psfrag{I}[cc][cc]{$I$}
      \psfrag{II}[cc][cc]{$II$}
      \psfrag{III}[cc][cc]{$III$}
      \psfrag{Site, n}[tt][tc]{Site, $n$}
      \psfrag{Time, t}[Bc][Bc]{Time, $t$}
    \epsfig{file=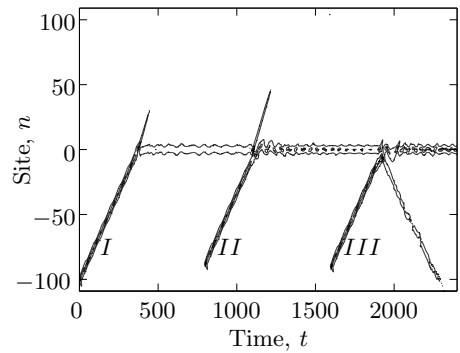, width=6 cm, angle=0} } 
  \caption{Trapping of breathers $I$, $II$ and $III$ at $n=0$. Contour plot 
  	for the evolution of the Hamiltonian density, $\mathcal{H}_n$, with 5 
	equidistant lines $H=0.005 \ldots 0.05$. $N_T=301$, $J_0=0.5$, 
	$v=0.2$, $k=0.2$, $A=0.5$. $I:$ $\nu=-100$ at $t=0$, $II:$ $\nu=-87$ 
	at $t=800$, $III:$ $\nu=-89$ $t=1600$.}
  \label{fig:contour_sum}
\end{figure}

In Fig.~\ref{fig:contour_sum} we show the trapping of multiple breathers. A 
first Gaussian pulse ($I$) is launched at site $\nu=-100$ at $t=0$. At $t=800$ 
both the displacements and the velocities are set to zero, $u_n(800)=0$ and 
$\dot{u}_n(800)=0$, outside the bent region, $|n|>15$, to remove radiation. 
This ``cleaned'' chain is now used as an initial condition for a new 
simulation, in which we add a second identical Gaussian pulse ($II$) launched 
at site $\nu=-87$. Like in other systems the interaction between two 
breathers, or a breather and an impurity, depends strongly on the relative 
phase. We choose $\nu=-87$ for the launching of this second pulse to obtain 
maximal trapping. Using the same procedure, a third identical Gaussian pulse 
($III$) is launched at $t=1600$, now at $\nu=-89$.
As seen in Fig.~\ref{fig:contour_sum} we essentially succeed in trapping 3 
breathers at the tip of the wedge chain. Some energy transmission is observed 
when breathers $I$ and $II$ are trapped, while reflection occurs at the 
trapping of breather $III$. By successively handling the initial conditions as 
described above, we avoid radiation which, when reflected at the boundaries, 
distorts the numerical simulations.\par 
\begin{figure}[!htb]
      \psfrag{y5}[cr][cr]{$0.2$}
      \psfrag{y4}[cr][cr]{$$}
      \psfrag{y3}[cr][cr]{$0.1$}
      \psfrag{y2}[cr][cr]{$$}
      \psfrag{y1}[br][cr]{$0$}
      \psfrag{x1}[cc][cc]{$0$}
      \psfrag{x2}[cc][cc]{$500$}
      \psfrag{x3}[cc][cc]{$1000$}
      \psfrag{x4}[cc][cc]{$1500$}
      \psfrag{x5}[cc][cc]{$2000$}
      \psfrag{Hc}[tc][tc]{Central energy, $H_c$}
      \psfrag{Time, t}[Bc][Bc]{Time, $t$}
  \centerline{\epsfig{file=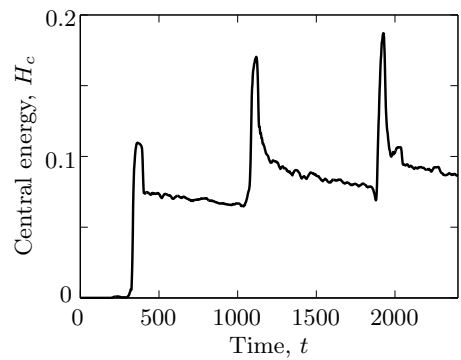, width=6 cm, angle=0} } 
  \caption{Central energy, $H_{c}$, versus time corresponding to 
  	Fig.~\ref{fig:contour_sum}.}
  \label{fig:Hcen_sum}
\end{figure}

The corresponding energy evolution for the central sites is shown in 
Fig.~\ref{fig:Hcen_sum}. The ability of the system to trap energy at the 
bending region is evident, even though more radiation is observed as the 
number of trapped breathers increases. As noted also in connection with 
Fig.~\ref{fig:Hcen}(b), the first incident breather, $I$, loses about 36\% of 
the total energy before trapping. For the following breathers, $II$ and $III$, 
both of the corresponding losses amount to 50\%. Thus the possibility for 
trapping more energy at the chain bend by additional incoming breathers may 
seem exhausted due to an effective saturation.


\section{\label{sec:random}Funneling}
The trapping of breathers observed in the previous Section suggests that the 
bend may funnel energy from the surrounding region. In order to study this in 
detail, we now explore the dynamics of the chain in the case of random initial 
conditions. 500 different realizations with zero displacement, $u_n=0$, and 
velocities normally distributed with zero mean, $<\!\!\dot{u}_n\!\!>=0$ and 
standard deviation $\sigma_{\dot{u}_n}=0.17$ are used, corresponding to a 
Hamiltonian $H \approx 1.41$.\par
\begin{figure}[h] 
      \psfrag{y4}[tr][tr]{$1$}
      \psfrag{y3}[tr][tr]{$0.5$}
      \psfrag{y2}[tr][tr]{$0$}
      \psfrag{y1}[tr][tr]{$-0.5$}
      \psfrag{x1}[cc][cc]{$0$}
      \psfrag{x2}[cc][cc]{$50$}
      \psfrag{x3}[cc][cc]{$100$}
      \psfrag{x4}[cc][cc]{$150$}
      \psfrag{Time, t}[cc][Bc]{Time, $t$}
      \psfrag{Displacement}[bc][tc]{Displacement, $u_0$}
  \centerline{\epsfig{file=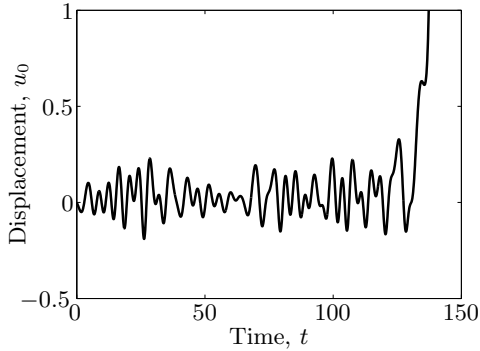, width=6 cm, angle=0} } 
  \caption{Displacement at center site, $u_0$, versus time on bent chain with 
  	$N_T=99$ and $\theta=95^{\circ}$. Initial displacements: $u_n=0$ for 
	all $n$. Random initial velocities generated with normal distribution: 
	$<\!\!\dot{u}_n\!\!>=0$ and $\sigma_{\dot{u}_n}=0.17$. H=1.41.}
  \label{fig:site_open}
\end{figure}

 
Random initial disturbances may create nonlinear localized excitations, 
interacting with each other and with the effective inhomogeneity caused by the 
bend \cite{PD 119, PRE 60, Muto, PRE 47/1, PD 57}. In \cite{Muto}, the number 
of generated solitons were found to depend on the temperature of the system, 
$T$, by the power $1/3$. Here we find that collision of the nonlinear 
excitations may result in exponential growth of the oscillation amplitude at 
the collision site. We observe this phenomenon in Fig.~\ref{fig:site_open} in 
a bend chain. Here, a sudden increase of the amplitude of the center site, 
$u_0$, occurs after about 130 time units of bounded oscillations. 
This unbounded growth of amplitude implies energy localization at the center 
site.\par
\begin{figure}[h] 
      \psfrag{y4}[tr][tr]{$30$}
      \psfrag{y3}[tr][tr]{$20$}
      \psfrag{y2}[tr][tr]{$10$}
      \psfrag{y1}[tr][tr]{$0$}
      \psfrag{x1}[cc][cc]{$-40$}
      \psfrag{x2}[cc][cc]{$-20$}
      \psfrag{x3}[cc][cc]{$0$}
      \psfrag{x4}[cc][cc]{$20$}
      \psfrag{x5}[cc][cc]{$40$}
      \psfrag{Occurrences}[tc][tc]{Occurrences}
      \psfrag{Site, n}[Bc][Bc]{Site, $n$}
  \centerline{\epsfig{file=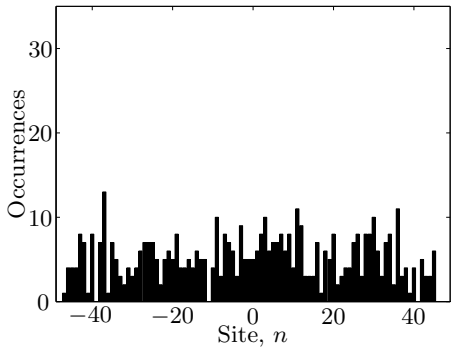, width=6 cm, angle=0} }  
  \caption{Straight chain, $\theta=180^{\circ}$, $N_T=99$. Occurrence of 
  	amplitudes above threshold, $u_n > 10$, versus site $n$, until max. 
	$t=10.000$. Initial displacements: $u_n=0$ for all $n$. 500 random 
	initial velocities with normal distribution: $<\!\!\dot{u}_n\!\!>=0$ 
	and $\sigma_{\dot{u}_n}=0.17$, H=1.41.}
  \label{fig:hist_theta180_017}
\end{figure}

Our results for a straight chain are depicted in 
Fig.~\ref{fig:hist_theta180_017} showing a histogram of the occurrences of 
sites with a displacement above the threshold value $u_n=10$, corresponding to 
the value for DNA opening used in \cite{JCP 122}. The simulations were 
discontinued when this threshold value was exceeded. If this event did not 
occur within 10.000 time units, a ``no--occurrence'' was registered. 
``No--occurrence'' happened in 24 out of 500 simulations. The threshold 
transgressions is seen to be uniformly distributed along the chain.\par
\begin{figure}[h] 
      \psfrag{y4}[tr][tr]{$30$}
      \psfrag{y3}[tr][tr]{$20$}
      \psfrag{y2}[tr][tr]{$10$}
      \psfrag{y1}[tr][tr]{$0$}
      \psfrag{x1}[cc][cc]{$-40$}
      \psfrag{x2}[cc][cc]{$-20$}
      \psfrag{x3}[cc][cc]{$0$}
      \psfrag{x4}[cc][cc]{$20$}
      \psfrag{x5}[cc][cc]{$40$}
      \psfrag{Occurrences}[tc][tc]{Occurrences}
      \psfrag{Site, n}[Bc][Bc]{Site, $n$}
  \centerline{\epsfig{file=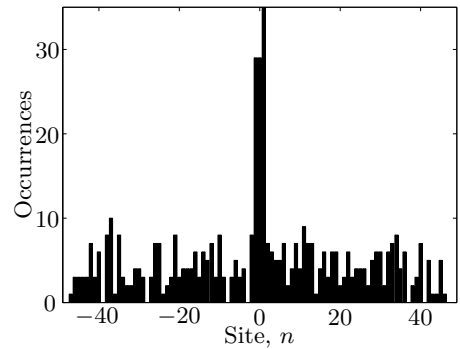, width=6 cm, angle=0} } 
  \caption{Wedge chain, $\theta=95^{\circ}$, $N_T=99$. Occurrence of 
  	amplitudes above threshold, $u_n > 10$, versus site $n$, until max. 
	$t=10.000$. Initial displacements: $u_n=0$ for all $n$. 500 random 
	initial velocities with normal distribution: $<\!\!\dot{u}_n\!\!>=0$ 
	and $\sigma_{\dot{u}_n}=0.17$, H=1.41.}
  \label{fig:hist_theta95_017}
\end{figure}

On a wedge chain with bending angle $\theta=95^{\circ}$, identical initial 
conditions gives the remarkably different result shown in 
Fig.~\ref{fig:hist_theta95_017}. Here, 93 (out of 491) threshold 
transgressions occurs in the center region $-1 \leq n \leq 1$. Only 9 
``no--occurrences'' were registered. Thus energy localization, implied by 
unbounded growth of amplitude, is observed in the vicinity of the tip of the 
wedge which therefore acts as an energy funnel.\par
For lower standard deviations, $\sigma_{\dot{u}_n}$, corresponding more 
realistically to physiological temperatures, a smaller number of simulations 
produce threshold transgressions, within 10.000 time units.
Thus, for $\sigma_{\dot{u}_n}=0.116$, $H \approx 0.7$, corresponding to the 
temperature $T \approx 37^{\circ}C$, only about 1\% of the 500 simulations led 
to threshold transgression. The few transgressions that did occur, were found 
in the vicinity of the tip of the wedge.\par
%

\section{\label{sec:conclu}Conclusion}
On a bent chain of Morse oscillators we find that moving discrete breathers 
may be trapped at a bending point in the presence of dipole--dipole--like 
longe--range interaction. Thus the role of the geometry for the dynamics is 
analogous to that of an inhomogeneity. At the bending point, several incident 
discrete breathers may be trapped. However, energy is lost to radiation and a 
saturation effect seems to limit the total trapped energy in the vicinity of 
a given bending point.\par
For random initial conditions modelling thermal fluctuations, the tendency to 
unbounded amplitude growth in the vicinity of the bending point is 
substantially amplified. Thus energy localization is implied in this region 
which therefore acts as an energy funnel.\par
The use of a nonlinear potential is crucial for obtaining the energy 
funnneling effect in our model, since no amplitude growth is observed in a 
linear approximation. The plateau of the characteristic Morse potential 
allows for the breaking of the hydrogen bonds, {\it e.g.}, in the molecule to 
be modelled. In contrast, a linear approximation with a parabolic potential 
produces too powerful an attraction for this effect to take place and would 
therefore not be an adequate description of chemical bonds.\par
We also note that the attracting nature of the long--range interaction is 
crucial for the occurrence of amplitude growth. Ongoing work on this effect 
uses a more accurate dipole--dipole interaction term and includes the effect 
of the twisting of the dipoles occurring along the strands of DNA molecules, 
{\it e.g.}\par
%

\section*{Acknowledgements}
The authors wish to thank S.F. Mingaleev for helpful and inspiring 
contributions. Yu.B.G. thanks Informatics and Mathematical Modelling, 
Technical University of Denmark for hospitality. O.B. acknowledges support 
from the Danish Technical Research Council (Grant No.~26-00-0355). 
P.V.L. thanks the Oticon Fondation for a study grant. The work is supported 
by LOCNET Project No. HPRN-CT-1999-00163. 

\pagebreak


\begin{thebibliography}{99}
%
\bibitem{Davy} A.S. Davydov: 
  {\it ``Solitons in Molecular Systems''} 
  (D. Reidel, Dordrecht, 1985).
%
\bibitem{RMP 60} A.J. Heeger, S. Kivelson, J.R. Schrieffer and W.P. Su, 
	Rev. Mod. Phys {\bf 60}, 781 (1988).
%
\bibitem{PR 217} A.C. Scott, 
	Phys. Rep. {\bf 217}, 1 (1992).
%
\bibitem{NLinBio} M. Peyrard (Ed.): 
	{\it "Nonlinear Excitations in Biomolecules"} 
	(Springer, Les Ulis, 1995).
%
\bibitem{PR 295} S. Flach and C.R. Willis, 
	Phys. Rep. {\bf 295}, 181 (1998).
%
\bibitem{Yaku} L.V. Yakushevich: 
  {\it ``Nonlinear Physics of DNA''} (John Wiley \& Sons, New York, 1998).
%
\bibitem{PRE 67} I. Bena, A. Saxena, G.P. Tsironis, M. Iba\~nes and 
		J.M. Sancho, 
	Phys. Rev. E {\bf 67}, 037601 (2003).
%
\bibitem{PRE 55/4} K. Forinash, T. Cretegny and M. Peyrard, 
	Phys. Rev. E {\bf 55}, 4740 (1997).
%
\bibitem{Muto}  
	V. Muto, A.C. Scott and P.L. Christiansen, 
		Phys. Lett. A {\bf 136}, 33 (1989).\\
	V. Muto, A.C. Scott and P.L. Christiansen, 
		Physica D, {\bf 44}, 75 (1990).\\
	V. Muto, 
		Nanobiology {\bf 1}, 325 (1992).
%
\bibitem{PRE 53/1} J.J.-L. Ting and M. Peyrard, 
	Phys. Rev. E {\bf 53}, 1011 (1996).
%
\bibitem{JPA 35_10519} J. Cuevas, F. Palmero, J.F.R. Archilla and F.R. Romero, 
	J. Phys. A {\bf 35}, 10519 (2002).
%
\bibitem{PRE 49} K. Forinash, M. Peyrard and B. Malomed, 
	Phys. Rev. E {\bf 49}, 3400 (1994).
%
\bibitem{PLA 154} H. Feddersen, 
	Phys. Lett. A {\bf 154}, 391 (1991).
%
\bibitem{Bang/Peyrard} 
	O. Bang and M. Peyrard, 
		Physica D {\bf 81}, 9 (1995).\\
	O. Bang and M. Peyrard, Phys. 
		Rev. E {\bf 53}, 4, 4143 (1996).%
%
\bibitem{JBP 25} S.F. Mingaleev, P.L. Christiansen, Yu.B. Gaididei, 
		M. Johansson and K.{\O}. Rasmussen, 
	J. Biol. Phys. {\bf 25}, 41 (1999).
%
\bibitem{PRL 70} T. Dauxois and M. Peyrard, 
	Phys. Rev. Lett. {\bf 70}, 3935 (1993).
%
\bibitem{PRE 51} F. Zhang, M.A. Collins and Yu.S. Kivshar, 
	Phys. Rev. E {\bf 51}, 3774 (1995).
%
\bibitem{PLA 249} L. Cruzeiro--Hansson, 
	Phys. Lett. A {\bf 249}, 465 (1998).
%
\bibitem{PD 163} J.Cuevas, J.F.R. Archilla, Yu.B. Gaididei and F.R. Romero, 
	Physica D {\bf 163}, 106 (2002).
%
\bibitem{PLA 253} M. Barbi, S. Cocco and M. Peyrard, 
	Phys. Lett. A {\bf 253}, pp. 358--369 (1999).
%
\bibitem{JBP 24} M. Barbi, S. Cocco, M. Peyrard and S. Ruffo, 
	J. Biol. Phys. {\bf 24}, 97 (1999).
%
\bibitem{EPL 59} S.F.M. Mingaleev, Yu.B. Gaididei, P.L. Christiansen and 
		Yu.S. Kivshar, 
	Euro. Phys. Lett. {\bf 59}, 403 (2002).
%
\bibitem{JPA 34_8465} R. Reigada, J.M. Sancho, M.I. Iba\~nes and 
		G.P. Tsironis, 
	J. Phys. A {\bf 34}, 8465 (2001).
%
\bibitem{Cond 13} P.L. Christiansen, Yu.B. Gaididei and S.F. Mingaleev, 
	J. Phys. Cond. Matt. {\bf 13}, 1181 (2001).
%
\bibitem{JPA 35_8885} J.F.R. Archilla, Yu.B. Gaididei, P.L. Christiansen and 
		J. Cuevas, 
	J. Phys. A {\bf 35}, 8885 (2002).
%
\bibitem{PRE 66} B. S\'anchez-Rey, J.F.R. Archilla, F. Palmero and F.R. Romero, 
	Phys. Rev. E {\bf 66}, 017601 (2002).
%
\bibitem{JPA 34_6363} J.F.R. Archilla, P.L. Christiansen, S.F. Mingaleev and 
		Yu.B. Gaididei, 
	J. Phys. A {\bf 34}, 6363 (2001).
%
\bibitem{PRE 62} Yu.B. Gaididei, S.F. Mingaleev and P.L. Christiansen, 
	Phys. Rev. E {\bf 62}, R53 (2001).
%
\bibitem{PLA 299} J. Cuevas, F. Palmero, J.F.R. Archilla and F.R. Romero, 
	Phys. Lett. A {\bf 299}, 221 (2002).
%
\bibitem{PD 57} T. Dauxois, M. Peyrard and C.R. Willis, 
	Physica D {\bf 57}, 267 (1992).
%
\bibitem{Saenger} W. Saenger: 
  {\it ``Principles of Nucleic Acid Structure''} 
  (Springer Verlag, New York, 1984).
%
\bibitem{Und DNA} C.R. Calladine and H.R. Drew: 
  {\it ``Understanding DNA''} (Academic Press, London, 2002).
%
\bibitem{PRL 62} M. Peyrard and A.R. Bishop, 
	Phys. Rev. Lett. {\bf 62}, 2755 (1989).
%
\bibitem{PRE 47/1} T. Dauxois, M. Peyrard and A.R. Bishop, 
	Phys. Rev. E {\bf 47}, 684 (1993).
%
\bibitem{PRE 60} G.P. Tsironis, A.R. Bishop, A.V. Savin and A.V. Yolotaryuk, 
	Phys. Rev. E {\bf 60}, 6610 (1999).
%
\bibitem{PD 119} M. Peyrard, 
	Physica D {\bf 160}, 184 (1998).
%
\bibitem{PRE 63} A.A. Sukhorukov, Yu.S. Kivshar, O. Bang, J.J. Rasmussen and 
		P.L. Christiansen, 
	Phys. Rev. E {\bf 63}, 036601 (2001).
%
\bibitem{PVL} P.V. Larsen, IMM-THESIS-2002-32, 
	Technical University of Denmark, Lyngby (2002).
%
\bibitem{JCP 122} S. Cocco and R. Monasson, 
	J. Chem. Phys. {\bf 122}, 10017 (2000).
%
\end{thebibliography}
\end{document}